\documentclass[prb,preprint,showpacs,preprintnumbers,amsmath,amssymb]{revtex4}
\usepackage{bm}
\usepackage{epsfig}

\begin{document}
\title{Electric-field driven long-lived spin excitations on a cylindrical surface with
spin-orbit interaction}
\author{P. Kleinert}
\email[]{kl@pdi-berlin.de}
\affiliation{Paul-Drude-Intitut f\"ur Festk\"orperelektronik,
Hausvogteiplatz 5-7, 10117 Berlin, Germany}
\date{\today}
\begin{abstract}
Based on quantum-kinetic equations, coupled spin-charge drift-diffusion equations
are derived for a two-dimensional electron gas on a cylindrical surface. Besides the
Rashba and Dresselhaus spin-orbit interaction, the elastic scattering on impurities,
and a constant electric field are taken into account. From the solution of the
drift-diffusion equations, a long-lived spin excitation is identified for spins
coupled to the Rashba term on a cylinder with a given radius. The electric-field
driven weakly damped spin waves are manifest in the components of the magnetization
and have the potential for non-ballistic spin-device applications.
\end{abstract}


\maketitle

\section{Introduction}
In the emerging field of spintronics, a main issue is the avoidance of spin randomization while
stimulating controlled rotations of spins in a spin field effect transistor. Therefore, the
recent proposal \cite{PRL_146801} for a spintronic device, which operates in the non-ballistic
regime, received considerable interest. According to this design, spin relaxation is suppressed,
when the Rashba and Dresselhaus spin-orbit interaction (SOI) constants are tuned by an external
gate voltage so that their couplings become equal. In fact, the suppression is a consequence of
an exact spin-rotation symmetry, \cite{PRL_236601,PRL_076604W} which leads to a strong
anisotropy of the in-plane spin-dephasing time. At the presence of an in-plane electric field,
the persistent spin helix is converted into a field-dependent internal eigenmode.\cite{PRB_205326}
In close analogy to space-charge waves in crystals, these field mediated spin excitations can
be probed by optical grating techniques.\cite{PRL_076604W} Both spin and charge pattern, which
are generated by polarized laser beams, provide the required wave vector for the excitation of internal
eigenmodes. Suppressed spin-relaxation occurs not only in (001) GaAs/Al$_x$Ga$_{1-x}$As quantum
wells with balanced Rashba and Dresselhaus SOI strengths but also in (110) quantum wells with
Dresselhaus coupling.\cite{PRL_4196}

Most activities in the field of spintronics that are based on the Rashba and Dresselhaus SOI refer
to a plane two-dimensional electron gas (2DEG), which is confined by a semiconductor quantum well.
However, the diversity of spin-related phenomena markedly increases for different geometries of
the 2DEG. In dependence on the curvature of the surface, in which the 2DEG resides, additional
contributions to the SOI appear that may lead to new spin effects. Examples of current interest
provide microtubes fabricated by exploiting the self-rolling mechanism of strained bilayers.
These rolled-up structures exhibit pronounced optical resonances \cite{PRL_077403} arising from
micron-sized cylindrical resonators or give rise to novel magnetoresistance oscillations, which
were observed in the ballistic transport of electrons on cylindrical surfaces.\cite{Frie_2008}
For non-ballistic spintronic device applications, the prediction of a conserved spin component,
which arises when the Rashba coupling constant $\gamma_1$ equals the quantity $\hbar/2m^{*}R$
(with $R$ being the radius of the cylinder and $m^{*}$ the effective mass of the 2DEG) is most interesting.\cite{Trushin_2008} The identification of this novel long-lived spin mode by
fabricated curved samples seems to be feasible with the present-day technology. \cite{Nature_410,APL_212113,PRB_205309,PRB_045347} It is the aim of this paper to study the
field dependence of the predicted spin helix by a systematic consideration of electric-field
mediated eigenmodes of a spin-charge coupled 2DEG that is confined to a circular cylinder.
Both Rashba and Dresselhaus SOIs as well as spin-independent elastic impurity scattering are
taken into account.

\section{Basic theory}
While in classical mechanics the restricted motion of particles on curved surfaces is
unambiguously described by equations of motion, the quantum-mechanical study of curved
systems starts from two different perspectives. In the first, widely used method, the
three-dimensional Schr\"odinger equation is converted to its two-dimensional counterpart
by an appropriate confining procedure.\cite{PRA_23_1982} This approach naturally accounts
for the fact that the curved structures are embedded in a three-dimensional space, in
which electric and magnetic fields could be present. The second alternative description of the carrier
dynamics on curved samples completely rests on a two-dimensional model.\cite{deWitt} In our study of SOI on
the surface of a cylinder, we apply the widely accepted first approach, which was already
used for studying spin effects on curved surfaces.\cite{Magarill_96,Chaplik_98,Magarill_98,PRB_085330}
The second-quantized version of the Hamiltonian has the form\cite{Magarill_98}
\begin{eqnarray}
H_0&=&\int\limits_{0}^{\infty}\frac{d\varphi}{2\pi}\sum\limits_{k_z}
\biggl\{
\sum\limits_{s}a_{k_zs}^{\dag}(\varphi)
\left[\frac{\hbar^2k_z^2}{2m^{*}}+\frac{\widehat{p}_{\varphi}^2}{2m^{*}} \right]
a_{k_zs}(\varphi)\label{H0}\\
&+&\gamma_1\sum\limits_{s,s^{\prime}}a_{k_zs}^{\dag}(\varphi)
\left[\sigma_{ss^{\prime}}^{z}\widehat{p}_{\varphi}-\hbar k_z\Sigma_{ss^{\prime}} \right]
a_{k_zs^{\prime}}(\varphi)\nonumber\\
&+&\gamma_2\sum\limits_{s,s^{\prime}}a_{k_zs}^{\dag}(\varphi)
\left[\frac{1}{2}\left(\Sigma_{ss^{\prime}}\widehat{p}_{\varphi}+\widehat{p}_{\varphi}\Sigma_{ss^{\prime}} \right) -\hbar k_z\sigma_{ss^{\prime}}^{z}\right]
a_{k_zs^{\prime}}(\varphi)\nonumber
\biggl\},
\end{eqnarray}
in which Rashba and Dresselhaus contributions appear with the coupling constants $\gamma_1$ and $\gamma_2$,
respectively. The creation [$a_{k_zs}^{\dag}(\varphi)$] and annihilation [$a_{k_zs}(\varphi)$]
operators depend on the spin index $s$, the wave vector component $k_z$ along the axis of the
cylinder, and the angle $\varphi$. The SOI terms include the Pauli matrices ${\bm \sigma}$, the
transverse momentum operator $\widehat{p}_{\varphi}$, and a matrix $\widehat{\Sigma}$ that introduces
off-diagonal elements with respect to the spin index. These quantities are defined by
\begin{equation}
\widehat{p}_{\varphi}=-\frac{i\hbar}{R}\frac{\partial}{\partial \varphi},\quad
\widehat{\Sigma}=\left(
                   \begin{array}{cc}
                     0 & -ie^{-i\varphi} \\
                     ie^{i\varphi} & 0 \\
                   \end{array}
                 \right).
\end{equation}
The periodic boundary conditions on the cylinder surface are accounted for by a discrete Fourier
transformation, which is applied in the form
\begin{equation}
a_{k_z \uparrow}(\varphi)=\sum\limits_{m=-\infty}^{\infty}e^{im\varphi}a_{k_zm\uparrow},\quad
a_{k_z \downarrow}(\varphi)=e^{i\varphi}\sum\limits_{m=-\infty}^{\infty}e^{im\varphi}a_{k_zm\downarrow}.
\end{equation}
By this transformation, the projection of the total angular momentum on the cylinder axis appears
and the Hamiltonian simplifies considerably. In addition to the SOI, both elastic scattering on
impurities with the short-range coupling strength $U$ and an external electric field ${\bm E}$
(applied along the cylinder axis) are taken into account. As it is assumed throughout the paper
that the radius $R$ of the cylinder is much larger than the lattice constant, we introduce the
electron momentum vector
\begin{equation}
{\bm k}=(k_{\varphi},k_z,0),\quad k_{\varphi}=\left(m+\frac{1}{2}\right)/R,
\end{equation}
in order to express the Hamiltonian in a form that is very similar to the case of planar geometry. We obtain
\begin{eqnarray}
H&=&\sum\limits_{{\bm k},s}\varepsilon({\bm k})a^{\dag}_{{\bm k}s}a_{{\bm k}s}
+\sum\limits_{{\bm k}}\sum\limits_{s,s^{\prime}}(\hbar{\bm \omega}_{1}({\bm k})
\cdot{\bm \sigma}_{ss^{\prime}})a^{\dag}_{{\bm k}s}a_{{\bm k}s^{\prime}}\label{Hges}\\
&+&U\sum\limits_{{\bm k},{\bm k}^{\prime}}\sum\limits_{s}
a^{\dag}_{{\bm k}s}a_{{\bm k}^{\prime}s}
-ie{\bm E}\cdot\left. \sum\limits_{{\bm k},s}\nabla_{{\bm\kappa}}
a^{\dag}_{{\bm k}-\frac{{\bm \kappa}}{2}s}a_{{\bm k}+\frac{{\bm \kappa}}{2}s}\right|_{{\bm \kappa}={\bm 0}},
\nonumber
\end{eqnarray}
with the parabolic dispersion relation
\begin{equation}
\varepsilon({\bm k})=\frac{\hbar^2{\bm k}^2}{2m^{*}}-\frac{\hbar}{2R}
\left(\gamma_1-\frac{\hbar}{4m^{*}R} \right).
\end{equation}
The main part of the SOI is included in the vector
\begin{equation}
{\bm\omega}_1({\bm k})=\left(0,-(\gamma_1k_z-\gamma_2k_{\varphi}),k_{\varphi}(\gamma_1-\frac{\hbar}{2m^{*}R})-\gamma_2k_z\right).
\end{equation}
The model description of spin-independent scattering on impurities in Eq.~(\ref{Hges}) has the advantage of simplicity,
permitting us an exact treatment of scattering in the Born approximation via the scattering time
$\tau$ defined by
\begin{equation}
\frac{1}{\tau}=\frac{2\pi U^2}{\hbar}\sum\limits_{{\bm k}^{\prime}}\delta
(\varepsilon({\bm k})-\varepsilon({\bm k}^{\prime})).
\end{equation}
All information about the spin-orbit coupled electron ensemble is provided by the spin-density matrix
\begin{equation}
f_{s^{\prime}}^{s}({\bm k},{\bm k}^{\prime}|t)=\langle a^{\dag}_{{\bm k}s}
a_{{\bm k}^{\prime}s^{\prime}}\rangle_{t},
\end{equation}
which is calculated from quantum-kinetic equations. A transparent physical interpretation
of final results is facilitated by considering the projected spin vector on a local trihedron. This
transformation is achieved by
\begin{equation}
{\bm f}=\sum\limits_{s,s^{\prime}}f_{s^{\prime}}^{s}{\bm S}_{ss^{\prime}},
\end{equation}
with the following transformation matrices
\begin{equation}
S^{\varphi}=\frac{1}{2}\left(
                         \begin{array}{cc}
                           0 & ie^{2i\varphi} \\
                           -ie^{-2i\varphi} & 0 \\
                         \end{array}
                       \right), \quad
S^{z}=\frac{1}{2}\left(
                         \begin{array}{cc}
                           1 & 0 \\
                           0 & -1 \\
                         \end{array}
                       \right), \quad
S^{r}=\frac{1}{2}\left(
                         \begin{array}{cc}
                           0 & e^{2i\varphi} \\
                           e^{-2i\varphi} & 0 \\
                         \end{array}
                       \right),
\end{equation}
which project to the cylinder axis ($S_z$), as well as to the tangential ($S_{\varphi}$) and normal ($S_{r}$)
directions. To proceed, the wave vectors are shifted according to ${\bm k}\rightarrow{\bm k}+{\bm\kappa}/2$
and ${\bm k}^{\prime}\rightarrow{\bm k}-{\bm\kappa}/2$ with $k_{\varphi}=(m+m^{\prime}+1)/2R$
and $\kappa_{\varphi}=(m-m^{\prime})/R$. The derivation of spin-charge coupled kinetic
equations is carried out by applying the same steps as in our previous study of the planar
geometry.\cite{PRB_165313} The final result is expressed by the kinetic equations
\begin{equation}
\frac{\partial}{\partial t}f({\bm k},{\bm \kappa}|t)
-\frac{i\hbar}{m^{*}}{\bm k}\cdot{\bm\kappa}f+i{\bm\omega}({\bm\kappa})\cdot{\bm f}
+\frac{e}{\hbar}{\bm E}\cdot\nabla_{\bm k}f=\frac{1}{\tau}(\overline{f}-f),
\label{kin1}
\end{equation}
\begin{eqnarray}
&&\frac{\partial}{\partial t}{\bm f}({\bm k},{\bm \kappa}|t)
-\frac{i\hbar}{m^{*}}({\bm k}\cdot{\bm\kappa}){\bm f}-2{\bm \omega}({\bm k})\times{\bm f}+i{\bm\omega}({\bm\kappa})f
+\frac{e}{\hbar}({\bm E}\cdot\nabla_{\bm k}){\bm f}\nonumber\\
&&=\frac{1}{\tau}(\overline{{\bm f}}-{\bm f})
-\frac{\hbar{\bm \omega}({\bm k})}{\tau}\frac{\partial}{\partial\varepsilon({\bm k})}\overline{f}
+\frac{1}{\tau}\frac{\partial}{\partial\varepsilon({\bm k})}\overline{\hbar{\bm\omega}({\bm k})f},
\label{kin2}
\end{eqnarray}
in which a new SOI vector appears given by
\begin{equation}
{\bm \omega}({\bm\kappa})=(\omega_{1y}({\bm\kappa})\sin(2\varphi),\omega_{1y}({\bm\kappa})\cos(2\varphi),
-\omega_{1z}({\bm\kappa})).
\end{equation}
On the right-hand side of Eq.~(\ref{kin2}), there remain scattering contributions that are proportional
to the SOI. These terms are necessary for a consistent treatment of a homogeneous 2DEG. The bar over
quantities in Eqs.~(\ref{kin1}) and (\ref{kin2}) indicates an integration over the polar angle $\alpha$
of the vector ${\bm k}=k(\cos\alpha,\sin\alpha,0)$. The kinetic Eqs.~(\ref{kin1}) and (\ref{kin2}) serve
as a starting point for various studies of spin effects on cylindrical surfaces. We mention only the
field-induced spin accumulation and the influence of the spin degree of freedom on the charge current
(as well as the appearance of a "{}spin current"{} on the cylinder).\cite{PRB_165313} In this paper,
we do not follow this interesting line of reasoning but look for a solution of Eqs.~(\ref{kin1}) and
(\ref{kin2}) in the long-wavelength and low-frequency drift-diffusion regime.

The envisaged macroscopic behavior of the coupled spin-charge system is established during an evolution
period, in which a nonequilibrium spin polarization and charge density still exist, whereas the energy of
particles already thermalized.\cite{PRB_075340} In this transport regime, the following separation ansatz
for the mean components $\overline{f}$ and $\overline{{\bm f}}$ is justified
\begin{equation}
\overline{f}({\bm k},{\bm\kappa}|t)=-F({\bm\kappa}|t)\frac{n^{\prime}(\varepsilon({\bm k}))}{dn/d\varepsilon_F},\quad
\overline{{\bm f}}({\bm k},{\bm\kappa}|t)=-{\bm F}({\bm\kappa}|t)\frac{n^{\prime}(\varepsilon({\bm k}))}{dn/d\varepsilon_F},
\end{equation}
where $n(\varepsilon({\bm k}))$ denotes the Fermi distribution function and $n=\int d\varepsilon\rho(\varepsilon)n(\varepsilon)$ is the equilibrium carrier density (with $\rho(\varepsilon)$
being the density of states). $n^{\prime}(\varepsilon({\bm k}))$ is a short-hand notation for
$d n/d\varepsilon({\bm k})$. Adopting this approximation also for the field contributions on the
left-hand side of Eqs.~(\ref{kin1}) and (\ref{kin2}), we obtain a set of linear equations for the
components of the spin-density matrix. The solution is expanded with respect to ${\bm\kappa}$ and
integrated over the angle $\alpha$. This procedure leads to coupled equations for the charge $F$
and spin ${\bm F}$ distribution functions, the solution of which is easily integrated over the energy
$\varepsilon({\bm k})$. The resulting coupled spin-charge drift-diffusion equations take the form
\begin{equation}
\left[\frac{\partial}{\partial t}-i\mu{\bm E}\cdot{\bm\kappa}+D{\bm\kappa}^2 \right]F+\frac{i}{\mu_B}{\bm\omega}({\bm\kappa})\cdot{\bm M}
+\frac{2im^{*}\tau}{\hbar\mu_B}\left([{\bm\Lambda}\times{\bm\omega}({\bm \kappa})]\cdot{\bm M} \right)=0,
\label{charge}
\end{equation}
\begin{eqnarray}
&&\left[\frac{\partial}{\partial t}-i\mu{\bm E}\cdot{\bm\kappa}+D{\bm\kappa}^2+\widehat{\Gamma} \right]{\bm M}
+\frac{e}{m^{*}c}{\bm M}\times{\bm H}_{eff}\nonumber\\
&&-\chi(\widehat{\Gamma} {\bm H}_{eff})\frac{F}{n}-2i\frac{m^{*}\mu}{c}[{\bm\Lambda}\times{\bm\omega}({\bm\kappa})]F={\bm G},
\label{spin}
\end{eqnarray}
for the charge density $F$ and the magnetization ${\bm M}=\mu_B{\bm F}$ with $\mu_B=e\hbar/2m^{*}c$ being the
Bohr magneton. The vector ${\bm G}$ on the right-hand side of Eq.~(\ref{spin}) accounts for the source of
an external spin generation. Furthermore, $D$ and $\mu$ denote the diffusion coefficient and the mobility that
are related to each other via the Einstein relation $\mu=eDn^{\prime}/n$. The Pauli susceptibility is given by
$\chi=\mu_Bn^{\prime}$. Scattering times that refer to various spin components are collected by the symmetric
matrix $\widehat{\Gamma}$, which is given by
\begin{equation}
\widehat{\Gamma}=\frac{4Dm^{*2}}{\hbar^2}
\left(
  \begin{array}{ccc}
    a_{11}^2+a_{12}^2+a_{31}^2+a_{32}^2 & -(a_{22}a_{32}+a_{21}a_{31}) & -(a_{11}a_{21}+a_{22}a_{12}) \\
    -(a_{22}a_{32}+a_{21}a_{31}) & a_{11}^2+a_{12}^2+a_{21}^2+a_{22}^2 & -(a_{12}a_{32}+a_{11}a_{31}) \\
    -(a_{11}a_{21}+a_{22}a_{12}) & -(a_{12}a_{32}+a_{11}a_{31}) & a_{21}^2+a_{22}^2+a_{31}^2+a_{32}^2 \\
  \end{array}
\right),
\end{equation}
where the quantities $a_{ij}$ are expressed by the spin-orbit coupling constants
\begin{equation}
a_{11}=\gamma_2\sin(2\varphi),\quad
a_{21}=\gamma_2\cos(2\varphi),\quad
a_{12}=-\gamma_1\sin(2\varphi),\quad
a_{22}=-\gamma_1\cos(2\varphi),
\end{equation}
\begin{equation}
a_{31}=-\left(\gamma_1-\frac{\hbar}{2m^{*}R} \right),\quad a_{32}=\gamma_2.
\end{equation}
The electric field is accounted for by the vector
\begin{equation}
{\bm\Lambda}=(a_{21}\mu E_{\varphi}+a_{22}\mu E_z,a_{31}\mu E_{\varphi}+a_{32}\mu E_z,
a_{11}\mu E_{\varphi}+a_{12}\mu E_z),
\end{equation}
from which an effective magnetic field
\begin{equation}
{\bm H}_{eff}=\frac{2m^{*2}c}{e\hbar}({\bm\Lambda}+2iD{\bm\omega}({\bm\kappa})),
\end{equation}
is derived that enters Eq.~(\ref{spin}) for the field-induced magnetization. The appearance of a magnetic
field ${\bm H}_{eff}$, which is solely due to the electric field, illustrates why there is a perfect
electric-field analog of the Hanle effect.\cite{Kalevich}

The drift-diffusion Eqs.~(\ref{charge}) and (\ref{spin}) for the charge density $F({\bm\kappa},t)$ and
magnetization ${\bm F}({\bm \kappa},t)$ provide the basis for the study of many spin-related phenomena
of a curved 2DEG in the drift-diffusion regime. Here, we shall focus on spin-related eigenmodes of
the system.

\section{Long-lived spin waves}
A solution of Eq.~(\ref{spin}) is searched for under the condition that the retroaction of spin on the
charge density can be neglected so that the carrier density is given by its equilibrium value ($F=n$).
Performing a Laplace transformation with respect to the time variable $t$ and introducing the abbreviations
\begin{equation}
{\bm M}^{\prime}={\bm M}-\chi{\bm H}_{eff},\quad \Sigma=s-i\mu{\bm E}\cdot{\bm\kappa}
+D{\bm\kappa}^2,
\end{equation}
with $s$ being the Laplace variable, Eq.~(\ref{spin}) is converted into the linear equations,
\begin{eqnarray}
&&\Sigma{\bm M}^{\prime}+\widehat{\Gamma}{\bm M}^{\prime}
+\frac{e}{m^{*}c}{\bm M}^{\prime}\times{\bm H}_{eff}\\
&&={\bm M}(0)+{\bm G}/s-2i\frac{m^{*}\mu}{c}{\bm\Lambda}\times{\bm\omega}({\bm\kappa})n
-\chi\Sigma{\bm H}_{eff},\nonumber
\end{eqnarray}
which are symbolically written as $\widehat{T}{\bm M}^{\prime}={\bm Q}$. Eigenmodes of the spin subsystem
are calculated from the zeros of the determinant of the matrix $\widehat{T}$. A simple but cumbersome
algebra leads to the result
\begin{equation}
{\rm det}\widehat{T}=\Sigma(\sigma^2+{\bm \omega}_{H}^2)+g_2\left(\sigma+\frac{(\mu{\bm E})^2}{D} \right),
\end{equation}
in which the short-hand notations ${\bm \omega}_{H}=(e/m^{*}c){\bm H}_{eff}$ and $\sigma=\Sigma+g_1$
are used. The coupling constants $g_1$ and $g_2$ are given by
\begin{equation}
g_1=2\frac{4Dm^{*2}}{\hbar^2}\left[\gamma_1^2+\gamma_2^2-
\frac{\hbar}{2m^{*}R}\left(\gamma_1-\frac{\hbar}{4m^{*}R} \right) \right],
\end{equation}
\begin{equation}
g_2=\left(\frac{4Dm^{*2}}{\hbar^2} \right)^2
\left[\gamma_2^2-\gamma_1\left(\gamma_1-\frac{\hbar}{2m^{*}R} \right) \right]^2.
\end{equation}
The cubic equation det~$\widehat{T}=0$ with respect to the Laplace variable $s$ has three solutions, which
give the dispersion relations of spin excitations. Most eigenmodes have a finite lifetime. However, there
is one long-lived spin excitation, whose damping completely disappears for a given wave number $\kappa_z$.
This mode appears for a model without any Dresselhaus SOI ($\gamma_2=0$), when the coupling constant
$\gamma_1$ matches the quantity $\hbar/2m^{*}R$. In this case, we obtain ($s\rightarrow i\omega$)
\begin{equation}
\omega_{1,2}=-\mu E_z\left(\kappa_z\pm K \right)
-iD\left(\kappa_z\pm K \right)^2,
\end{equation}
with $K=2m^{*}\gamma_1/\hbar$ being a wave number that is built from the Rashba spin-orbit coupling
constant $\gamma_1$. This soft mode becomes increasingly undamped in the limit
$\kappa_z\rightarrow K$. The persistent spin mode of this kind, which is a consequence of a new
spin-rotation symmetry, has no counterpart in the planar Rashba model and is a distinct feature
that solely appears on a cylinder surface.

In order to excite the persistent spin wave, a regular lattice of spin polarization $Q_r$
perpendicular to the cylinder surface is provided by laser pulses. For simplicity,
the spin generation is assumed to have the form
\begin{equation}
Q_r=\frac{Q_{r0}}{2}[\delta(\kappa_z-\kappa_0)+\delta(\kappa_z+\kappa_0)].
\end{equation}
Under the condition $Q_{\varphi}=Q_{z}=0$, the solution ${\bm M}=\chi{\bm H}_{eff}+\widehat{T}^{-1}{\bm Q}$ of
Eq.~(\ref{spin}) is expressed by
\begin{equation}
M_z=\frac{\mu E_z K}{\sigma^2+{\bm \omega}_{H}^2}Q_r,\quad M_r=\frac{\sigma}{\sigma^2+{\bm \omega}_{H}^2}Q_r.
\end{equation}
In the derivation of these equations, it was considered that the inverse Fourier transformation
with respect to $\kappa_{\varphi}$ leads to $\varphi=0$. The inverse Laplace transformation and
the integration over $k_z$ give for the non-vanishing components of the field-mediated magnetization
the final results
\begin{eqnarray}
M_r(z,t)=\frac{Q_{r0}}{2}&&\biggl\{
e^{-D(\kappa_0+K)^2t}\cos[\kappa_0z+\mu E_z(\kappa_0+K)t]\nonumber\\
&&+e^{-D(\kappa_0-K)^2t}\cos[\kappa_0z+\mu E_z(\kappa_0-K)t]\biggl\},
\end{eqnarray}
\begin{equation}
M_z(z,t)=M_z^{(-)}(z,t)-M_z^{(+)}(z,t),
\end{equation}
with
\begin{eqnarray}
M_z^{\pm}(z,t)&=&\frac{\mu E_z}{(\mu E_z)^2+(2D\kappa_0)^2}\frac{Q_{r0}}{2}
\biggl\{2D\kappa_0\cos[\kappa_0z+\mu E_z(\kappa_0\pm K)t]\nonumber\\
&-&\mu E_z\sin[\kappa_0z+\mu E_z(\kappa_0\pm K)t]\biggl\}e^{-D(\kappa_0\pm K)^2t}.
\end{eqnarray}
Both components $M_z$ and $M_r$ consist of a strongly and weakly damped oscillating term.
Under the resonance condition $\kappa_0=K$, the first mode quickly disappears, whereas the
second mode becomes completely undamped. A smooth dependence on the electric field
$E_z$ persists in the magnetization $M_z$ along the cylinder axis. A slight detuning of
the resonance, however, leads to the appearance of an electric-field driven spin wave,
the damping of which is extremely weak. The frequency of this long-lived spin excitation
is directly controlled by the applied electric field. The situation is similar to the
persistent spin helix of a planar 2DEG. Therefore, it is supposed that the robust spin
wave on a cylinder and its direct manipulation by an electric field has the potential
to be utilized in future spintronic device applications.

\section{Summary}
Nanostructures with a great variety of novel geometries like curved graphene systems and
rolled-up 2DEG are now experimentally available. Hence, the rigorous theoretical study
of the dynamics on such curved surfaces became a subject of recent interest. Especially spin effects
have been treated because the curvature of the surface gives rise to additional contributions
to the SOI. Consequently, the number of possible spin effects considerably increases in
nanostructures with curved geometries. This observation further stimulates activities in the
field of spintronics. A key point regarding spin-field-effect transistors is the exclusive manipulation
of spin by means of an electric field. Particularly attractive is the proposal for a device working
in the non-ballistic regime, where spin scattering in a planar 2DEG is suppressed due to a
spin-rotation symmetry.\cite{PRL_146801} A similar effect that occurs on the surface of a
cylinder was studied in the present paper. Based on quantum-kinetic equations for the spin-density
matrix, rigorous coupled spin-charge drift-diffusion equations were systematically derived
for a cylinder, whose radius is much larger than the lattice constant. From the solution of these
equations, the dispersion relation of field-dependent spin eigenmodes are identified. In general,
there are three damped spin excitations, the character of which is determined by the coupling
constants $\gamma_1$ and $\gamma_2$ of the Rashba and Dresselhaus SOI. For the pure Rashba
model ($\gamma_2=0$), a long-lived spin wave exists, when the radius $R$ of the cylinder
matches the condition $R=\hbar/2m^{*}\gamma_1$. This finding is of particular interest as an
applied electric field stimulates a nearly undamped spin wave. The excitation mechanism of
spin waves has the same character as the excitation of space-charge waves, which are normally
strongly damped. Nevertheless, space-charge waves in crystals were clearly demonstrated in
experiment. To my knowledge, completely undamped space-charge waves do not exist. Their
damping is reduced, however, in the regime of negative differential conductivity  due to
a negative Maxwellian relaxation time.\cite{Bryksin_Petrov} The complete disappearance of
the damping of an excitation is a novelty that occurs in special spin subsystems with a
$k$-linear SOI. The above mentioned peculiarity of the Rashba model on a cylindrical surface
has no counterpart in a planar 2DEG. Unfortunately, the experimental demonstration of this
effect is rendered more difficult because the huge internal strain within the tube breaks
the bulk inversion symmetry so that an appreciable Dresselhaus contribution to the SOI is
expected, which detunes the strong spin resonance. If this problem can be circumvented,
the long-lived field-mediated spin excitations on a cylinder have the potential to be utilized
in spintronic devices that work even in the non-ballistic regime.

\bibliographystyle{prsty}
\bibliography{abbrev,spin}


\end{document}